\documentclass[11pt,fleqn]{article}
\usepackage{amssymb}
\usepackage[latin1]{inputenc}
\usepackage{times}
\usepackage{mathptm}
\usepackage{epsfig}
\newlength{\picttext}
\newcommand{\Section}[1]
{\section{#1}\setcounter{equation}{0}}
\newcommand{\ket}[1]{|\hspace{0.1 ex}#1\rangle}
\newcommand{\bra}[1]{\langle#1|}
\begin{document}
\mathindent 0mm
\thispagestyle{empty} 
\begin{flushright} HD--TVP--98--04\end{flushright}
\vspace*{1cm}
\begin{center} 
{\Large Flow Equations for the H\'enon--Heiles Hamiltonian
\\ \vspace*{0.4cm}}
\vskip0.5cm
Daniel Cremers\footnote[1]{E--mail: cremers@itb.biologie.hu-berlin.de , new permanent address: \\Innovationskolleg Theoretische Biologie, Humbold Universität Berlin, Invalidenstra{\ss}e 43, D-10115 Berlin} 
and Andreas Mielke\footnote[2]
{E--mail: mielke@tphys.uni-heidelberg.de}\\ 
\vspace*{0.2cm}
Institut für Theoretische Physik,\\
Ruprecht--Karls--Universität,\\
Philosophenweg 19, \\
D-69120~Heidelberg, F.R.~Germany
\\[0.5cm]
\today
\end{center}
\begin{abstract}
  The H\'enon--Heiles Hamiltonian was introduced in 1964 \cite{HandH}
  as a mathematical model to describe the chaotic motion of stars in a
  galaxy.  By canonically transforming the classical Hamiltonian to a
  Birkhoff--Gustavson normalform Delos and Swimm obtained a discrete
  quantum mechanical energy spectrum.
  The aim of the present work is to first quantize the classical
  Hamiltonian and to then diagonalize it using different variants of
  flow equations, a method of continuous unitary transformations
  introduced by Wegner in 1994 \cite{Wegner94}.
  The results of the diagonalization via flow equations are comparable
  to those obtained by the classical transformation.  In the case of
  commensurate frequencies the transformation turns out to be less
  lengthy.  In addition, the dynamics of the quantum mechanical system
  are analyzed on the basis of the transformed observables.

\vspace{0.5cm}
\begin{tabular}{ll}
PACS-numbers: & 03.65.-w (Quantum mechanics),
\\ & 05.45.+b (Theory and models
of chaotic systems)
\end{tabular}

\vspace{0.5cm}


\end{abstract}

\Section{Introduction}

The H\'enon--Heiles Hamiltonian describes two one--dimensional
harmonic oscillators with a cubic interaction.  It is one of the
simplest Hamiltonians to display soft chaos in classical mechanics: by
increasing the total energy a transition from an integrable to an
ergodic system is induced.  Originally conceived to model the chaotic
motion of stars in a galaxy it later became an important milestone in
the development of the theory of chaos \cite{Gutz90}, partly because
of the conceptual simplicity of the model.

In order to investigate this continuous loss of integrability with
growing total energy Gustavson \cite{Gus66} transformed the classical
Hamiltonian by a series of canonical transformations to a
Birkhoff--Gustavson normal form \cite{Bir27}, 
which allowed him to construct an
additional constant of motion.  Thus he was able to analytically
reproduce Poincar\'e surfaces of section as obtained by numerical
integration.

The present work is based upon two publications (\cite{DS77} and
\cite{DS79}) in which Delos and Swimm used the classical Birkhoff
transformation \cite{Bir27}
as applied by Gustavson to analyze how the classically
chaotic behavior of the system is transformed into quantum mechanics.
The Birkhoff--Gustavson normal form is a power series in oscillator
Hamiltonians and thus allows a direct determination of a quantum
mechanical eigenvalue spectrum from the classical Hamiltonian.  For a
fixed set of parameters Delos and Swimm analytically calculated the
spectrum of the H\'enon--Heiles Hamiltonian which reproduced the
eigenvalues obtained by numerical diagonalization of finite matrices.

The main problem of the quantum mechanical H{\'e}non--Heiles
Hamiltonian is the fact that it is not bounded from below. It contains
a cubic potential. The classical motion discussed by Gustavson and by
Delos and Swimm corresponds to initial conditions near the local
minimum of the potential and to an energy that is below the saddle
point value of the potential. By these conditions the classical
motion is always restricted to a finite region.  In the corresponding
quantum problem, the particle will always tunnel through the barrier.
Therefore the eigenvalue spectrum calculated by Delos and Swimm is not
the real eigenvalue spectrum of the Hamiltonian. It describes
effective states that can be used to describe the dynamics near the
minimum of the potential and for times that are small compared to the
escape time.

The aim of the present paper is to first quantize the classical
Hamiltonian and to then diagonalize it using the method of flow
equations which was introduced by Wegner \cite{Wegner94} in 1994.  It
is clear that concerning the tunneling problem, the flow equations
have the same limitation as the quantization of the
Birkhoff--Gustavson normal form by Delos and Swimm. The bound states
and the eigenvalues obtained using flow equations allow only an
effective description for small times. One advantage of the flow
equations is that a correct and simple treatment of the system is
possible even if the two frequencies of the harmonic oscillators are
commensurate. This is not the case if Birkhoff--Gustavson normal form
is quantized.  Furthermore, the quantum mechanical treatment allows a 
description of the dynamics.

The structure of the paper is as follows.  The next section offers a
general introduction to the flow equation method.  In sections
\ref{cut} and \ref{iter} their application in two variants to the
H\'enon--Heiles model is treated.  Section \ref{results} contains some
results of the diagonalization: in a table the energy eigenvalues
obtained in the flow equation approach are compared with those
obtained by numerical matrix diagonalization for a fixed coupling
constant.  A graph shows the dependence of the calculated eigenvalues
upon the coupling strength.  In section \ref{commens} a case of
commensurate frequencies is treated and a similar table of eigenvalues
is obtained.  Section \ref{dynamics} offers a method to investigate
the dynamics of the quantum mechanical system, transition amplitudes
between the eigenstates of the uncoupled system are determined.  The
effect of growing coupling strength upon the transition amplitudes is
shown.  The last section contains a summary of the results of the
present work, gives a comparison to the work of Delos and Swimm and a
discussion of the limitations of the Birkhoff--Gustavson
transformation.

\Section{Flow Equations}
The method of flow equations consists in a continuous unitary
transformation of a given Hamilton operator $H$, which can be written
in differential form:
\begin{equation}
  \label{eq:flow}
  \frac{dH(\ell)}{d\ell} = [\eta(\ell),H(\ell)]\; .
\end{equation}
There are several possibilities to choose the antihermitean generator
$\eta$ so that $H(\infty)$ becomes diagonal. Wegner \cite{Wegner94}
proposed:
\begin{equation}
  \label{eq:eta}
 \eta(\ell) = [H_d(\ell),H(\ell)] = [H_d(\ell),H_r(\ell)] \; ,
\end{equation}
where $H_d$ and $H_r$ are the diagonal and the off-diagonal portions
of the Hamilton operator respectively. A detailed argument for the
usefulness of this choice of the generator can be found in
\cite{Wegner94}.  But the consistency can be easily verified, since in
the limit $\ell\rightarrow\infty$ as $H(\ell)$ becomes more diagonal
$\eta(\ell)$ will vanish and so will $\frac{dH(\ell)}{d\ell}$.

The flow equation method has been applied to various models (see e.g.
\cite{subohmic,correl,Lenz,am1,am2}). The
general behavior is such that terms that appear in $\eta$ 
(\ref{eq:eta}) will result in new terms in the transformed Hamilton
operator by (\ref{eq:flow}).  If this iterative process does
not result in a closed set of differential equations it can be forced
into such by defining an order for the appearing terms -- e.g. the
number of creation operators in them -- and neglecting all terms of
higher order.  This approach will be called {\em cut--off}.

A second approach to handle the system of differential equations
(\ref{eq:flow}) and (\ref{eq:eta}) for a given Hamilton operator $H =
H_d(0) + \lambda\, H_r(0)$ is to define the transformed Hamilton
operator as a power series in the coupling constant $\lambda$
\begin{equation}
\label{eq:sum}
  H(\ell) = \sum_{k=0}^\infty \lambda^k H_k(\ell)\; .
\end{equation}
The coefficients $H_k(\ell)$ can be determined iteratively and this
method shall be called {\em iteration}.  Both of these procedures --
the cut--off and the iteration -- will become more transparent as they
are applied to the H\'enon--Heiles Hamiltonian in the next two
sections.
\begin{figure}[!ht]
\begin{center}

  \epsfig{file=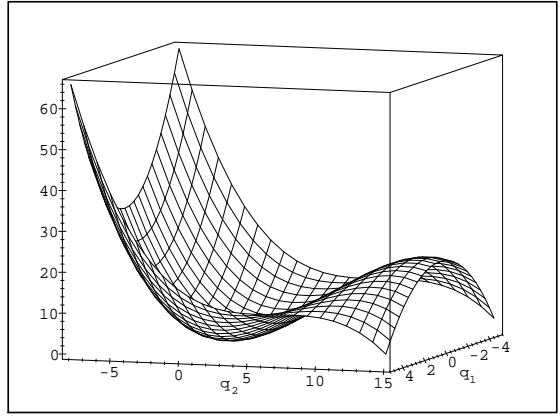,height=.5\textwidth,angle=270}
  \\[-4mm]
  \parbox[b]{.44\textwidth}{\caption[K]{\label{potential}
      H\'enon--Heiles potential (\ref{eq:hh_pq})\\ 
      ($\lambda\!=\!-0.1$, $n\!=\!0.1$, $w\!=\!1.3$, $ v\!=\!0.7$)}}
\end{center}
\end{figure}

\Section{The Cut--off Procedure}
\label{cut}
The H\'enon--Heiles Hamiltonian can be expressed as a function of two
spatial coordinates $q_1$ and $q_2$ and the two momenta $p_1$ and
$p_2$ (see figure \ref{potential}):
\begin{equation}
  \label{eq:hh_pq}
  H=\frac{1}{2}w(p_1^2+q_1^2)+\frac{1}{2}v(p_2^2+q_2^2)
  +\lambda q_2(q_1^2+n q_2^2)\; .
\end{equation}
Quantizing this classical Hamiltonian using the operators
\begin{equation}
\label{eq:a}
  a := \frac{1}{\sqrt{2}}(\hat{q}+i\hat{p}) \;, \quad a^{\dagger} :=
\frac{1}{\sqrt{2}}(\hat{q}-i\hat{p}) \;.
\end{equation}
will give the Hamilton operator
\begin{equation}
  \label{eq:hh}
  H=w\,a^\dagger a+v\,b^\dagger b+\lambda\,
(b^\dagger+b)\left((a^\dagger+a)^2+n\, (b^\dagger+b)^2\right)\; ,
\end{equation}
where the coupling constant $\lambda$ has been rescaled and the
constant term dropped.

This Hamilton operator is to be transformed into the quantum
mechanical equivalent of a Birkhoff normalform, a power series in
oscillator Hamiltonians:
\begin{equation}
  H \longrightarrow \sum_{k,m=0}^\infty \beta_{km}(a^\dagger a)^k
  (b^\dagger b)^m  \,= \sum_{k,m=0}^\infty w_{km}\;
  \,a^{\dagger k} a^k\:\: b^{\dagger m}b^m\,.
\end{equation}
The generalized frequencies $w_{km}$ are obtained from the transformed
Hamilton operator $H(\ell)$ in the limit $l\rightarrow\infty$.  A
consistent ansatz is $H(\ell)=H_d(\ell)+H_r(\ell)$ where:
\begin{eqnarray}
   H_d(\ell) & = & w(\ell) \:a^\dagger a + v(\ell) \: b^\dagger b \nonumber\\ &
& + \lambda^2 \,\bigg[\,w_{00}(\ell) + w_{20}(\ell)\:a^{\dagger 2} a^2 +
w_{02}(\ell)\:b^{\dagger 2} b^2 + w_{11}(\ell) \:a^\dagger a \:b^\dagger
b\,\bigg]  \quad \mbox{and}\nonumber\\
  H_r(\ell) & = &  \lambda \,\bigg[\,(a^{\dagger 2} + a^2)\,(b^\dagger + b)
\,x_1(\ell) \:+\:(a^{\dagger 2} - a^2)\,(b^\dagger - b)  \,x_2(\ell)
\nonumber\\
& &\quad\: +\:(a^\dagger a)\,(b^\dagger + b)  \,x_3(\ell) \:+\: (b^{\dagger 3}
+ b^3)  \,x_4(\ell)  \nonumber\\
& &\quad \:+\:(b^{\dagger 2}b + b^\dagger b^2)  \,x_5(\ell) \:+\:  (b^\dagger +
b)  \,x_6(\ell) \,\bigg] \,.
\end{eqnarray}
The $\ell$-dependent coefficients are determined from
(\ref{eq:flow}) and (\ref{eq:eta}) which combine to:
\begin{equation}
  \frac{dH}{d\ell} = \left[\left[H_d,H_r\right],H_d\right] +
\left[\left[H_d,H_r\right],H_r\right] \, .
\end{equation}
Neglecting all terms of third order in $\lambda$ or higher we obtain
the following set of differential equations by coefficient matching:
\begin{eqnarray}
  \label{eq:dequns}
 w_{00}^\prime & = & -4 x_1^2 v-8 x_1^2 w-8 x_1 x_2 v-16 x_1 x_2 w-36 x_4^2 v-2
x_6^2 v-4 x_2^2 v-8 x_2^2 w \nonumber\\
 w^\prime & = & -8 x_1^2 v-16 x_1^2 w-16 x_1 x_2 v-32 x_1 x_2 w-2 x_3^2 v-4 x_3
x_6 v-8 x_2^2 v-16 x_2^2 w \nonumber\\
 v^\prime & = & -16 x_1^2 w-16 x_1 x_2 v-108 x_4^2 v-4 x_5^2 v-8 x_5 x_6 v-16
x_2^2 w \nonumber\\
 w_{20}^\prime & = & -4 x_1^2 v-16 x_1 x_2 w-2 x_3^2 v-4 x_2^2 v \nonumber\\
 w_{11}^\prime & = & -32x_1^2 w-32 x_1 x_2 v-8 x_3 x_5 v-32 x_2^2 w \nonumber\\
 w_{02}^\prime & = & -54x_4^2 v-6 x_5^2 v \nonumber\\
 x_1^\prime & = & - x_1 v^2-4 x_1 w^2-4 x_2 v w \nonumber\\
 x_2^\prime & = & -4 x_1 v w- x_2 v^2-4 x_2 w^2 \nonumber\\
 x_3^\prime & = & - x_3 v^2 \nonumber\\
 x_4^\prime & = & -9 x_4 v^2 \nonumber\\
 x_5^\prime & = & - x_5 v^2 \nonumber\\
 x_6^\prime & = & - x_6 v^2
\end{eqnarray}
The equations for $x_3,\ldots,x_6$ are all of similar type, the
equations for $w_{ij}$ are uncoupled.  The system can therefore be
reduced to five equations.  It could not be solved analytically.  But
the asymptotic behavior for large $\ell$ is the following:

Assume that $w(\ell) \approx w(\infty) = w_\infty$ and $v(\ell)
\approx v(\infty) = v_\infty$.  Then the off-diagonal elements show an
exponential decay (except in the case of commensurate frequencies
$2w_\infty+v_\infty\!=\!0$):
\begin{eqnarray}
  \label{eq:asympt}
  x_1(\ell) & = & c_0 \exp\left(-(2w_\infty+v_\infty)^2\cdot
\ell\right)\nonumber\\
  x_2(\ell) & = & c_0 \exp\left(-(2w_\infty+v_\infty)^2\cdot
\ell\right)\nonumber\\
  x_3(\ell) & = & d_0 \exp\left(-v_\infty^2\cdot \ell\right)\, .
\end{eqnarray}

The diagonalized Hamilton operator was determined by numerical
integration of (\ref{eq:dequns}) using a Runge--Kutta
procedure.  For various sets of fixed parameters the approximation was
improved by extending the calculation to all terms up to fourth order
in $\lambda$.  The resulting set of differential equations for
coefficients $x_1,\ldots,x_{48},w,v$ and the coefficients $w_{ij}$ 
corresponding to the diagonal operators $(a^{\dagger i}
a^i)\,(b^{\dagger j} b^j)$ were also determined by numerical
integration.  An eigenvalue spectrum obtained from the transformed
Hamiltonian in the limit $\ell\rightarrow\infty$ was calculated for
different values of the coupling constant $\lambda$.  Results of these
calculations are presented in section \ref{results}.

\Section{The Iterative Procedure}
\label{iter}
A far more elegant way to solve the flow equation (\ref{eq:flow}) for
the H\'enon--Heiles Hamiltonian is an iterative calculation of the
$H_k(\ell)$ defined in (\ref{eq:sum}).  It avoids the
numerical integration applied in the previous section and allows a
better insight into the transformation mechanism, specifically the
behavior in the case of commensurate frequencies.  

The Hamiltonian (\ref{eq:hh}) is given in the form
\begin{equation}
H(0)=H_0(0) + \lambda\,H_1(0)\quad\mbox{with}\quad H_0(0)=w\,a^\dagger
a+v\,b^\dagger b\,.
\end{equation}
The transformed Hamiltonian is defined as a power series in $\lambda$
(\ref{eq:sum}).  In deviation from the original choice (\ref{eq:eta}) 
the generator is now defined as:
\begin{equation}
  \label{eq:eta2}
  \eta(\ell) := [H_0,H(\ell)\,] = \sum_{k=1}^\infty \lambda^k\,\eta_k(\ell) \,
, \qquad \mbox{where} \quad \eta_k = [H_0,H_k]\, .
\end{equation}
This choice of $\eta$ is in accordance with the classical Birkhoff
transformation as applied by Gustavson \cite{Gus66}.  It makes the
iterative calculation simpler, since commutation with $H_0$ will
reproduce a given operator term $a^{\dagger k} a^r b^{\dagger m} b^n$:
\begin{equation}
  \label{eq:repro}
  \left[\left[H_0\,,a^{\dagger k} a^r b^{\dagger m} b^n\right],H_0\right] =
-\epsilon_{krmn}\,a^{\dagger k} a^r b^{\dagger m} b^n\; ,
\end{equation}
where $\epsilon_{krmn}\!:=\!\left[(k-r)w+(m-n)v\right]^2$\,.
Inserting the power series (\ref{eq:sum}) for the transformed
Hamiltonian into the flow equation (\ref{eq:flow}) and comparing the
coefficients of the powers of $\lambda$ gives the differential
equations:
\begin{equation}
\label{eq:indiv}
  \frac{dH_n(\ell)}{d\ell} = \left[\left[H_0,H_n\right],H_0\right]\; +\;
\sum_{a+b=n \atop a,b\not=0} \left[\left[H_0,H_a\right],H_b\right]\quad
\forall\, n=0,1,2\ldots\,.
\end{equation}
It follows that $H_0=$ const. and $H_k(\ell),\, k=1,2\ldots$ can be
iteratively calculated: inserting the general ansatz
\begin{equation}
\label{eq:hn}
  H_n(\ell) = \sum_{k,r,m,n} \delta_{krmn}(\ell)\, a^{\dagger k} a^r b^{\dagger
m} b^n
\end{equation}
into (\ref{eq:indiv}) results in differential equations of
the form
\begin{equation}
  \label{eq:lambda}
  \frac{d}{d\ell}\delta_{krmn}(\ell) = -\epsilon_{krmn}\,\delta_{krmn}(\ell) +
\alpha_{krmn}(\ell)\quad \mbox{(no summation)}\,,
\end{equation}
where $\epsilon_{krmn}\ge 0$ was defined above. The function
$\alpha_{krmn}(\ell)$ can be shown to be a sum of terms of the form
\begin{equation}
\label{eq:typterm}
  c \ell^n \exp(-\gamma \ell)\qquad (\,c\!=\!\mbox{const.}
,\;\;\gamma\!>\!0,\;\; n\!=\!0,1,2\ldots\,)\,.
\end{equation}
The solution to (\ref{eq:lambda}) is
\begin{equation}
\label{eq:lambda2}
  \delta_{krmn}(\ell) = \exp(-\epsilon_{krmn}\cdot \ell)\,\int_0^\ell
d\ell^\prime \alpha_{krmn}(\ell^\prime)\exp(\epsilon_{krmn} \ell^\prime)\;.
\end{equation}
Ignoring for a moment the inhomogeneity $\alpha_{krmn}(\ell)$ we find
the following behavior: all terms will decay exponentially except for
two cases in which $\epsilon_{krmn}$ may vanish:
\begin{enumerate}
\item{The given operator term $a^{\dagger k} a^r b^{\dagger m} b^n$ is
    diagonal ($k=r$ and $m=n$)}.
\item{The frequencies $w$ and $v$ are commensurate.}
\end{enumerate}
It can be seen that the inhomogeneity $\alpha_{krmn}(\ell)$ will not
change this general behavior.  Thus the transformation can be
successfully performed in the case of incommensurate frequencies and
only diagonal terms will remain in the limit $\ell\rightarrow\infty$.  
In the case of commensurate frequencies off-diagonal operators
$a^{\dagger k} a^r b^{\dagger m} b^n$ may only remain for
$\epsilon_{krmn}=0$.

Since the number of terms in $H_k(\ell)$ grows rapidly with $k$ the
described procedure with all its algebraic manipulations was performed
by a computer program in the language C.  This way $H_0,\ldots,H_8$
were determined.  The results of these calculations will be presented
in the following section.

\Section{Results of the Diagonalization}
\label{results}
The two procedures described in the last two sections were applied to
the H\'enon-Heiles Hamiltonian (\ref{eq:hh_pq}) in a case of
incommensurate frequencies with the values $w\!=\!1.3$, $v\!=\!0.7$,
$\lambda\!=\!-0.1$, $n\!=\!0.1$.  The iteration procedure to 8th order
results in the following diagonal form (at $\lambda\!=\!-0.1$):
\begin{eqnarray}
  \label{eq:ordn}
  H & = & \;\;\, 2.20910\cdot 10^{-7}  a^{\dagger 5}\,a^5
+1.19855\cdot 10^{-6}  a^{\dagger 4}\,a^4b^\dagger b
+2.17973\cdot 10^{-6}  a^{\dagger 4}\,a^4
\nonumber\\ & & +1.54236\cdot 10^{-6}  a^{\dagger 3}\,a^3b^{\dagger 2}b^2
+4.12440\cdot 10^{-6}  a^{\dagger 3}\,a^3b^\dagger b
\nonumber\\ & & -8.65783\cdot 10^{-5}  a^{\dagger 3}\,a^3
+6.09731\cdot 10^{-8}  a^{\dagger 2}\,a^2b^{\dagger 3}b^3
\nonumber\\ & & -4.99568\cdot 10^{-6}  a^{\dagger 2}\,a^2b^{\dagger 2}b^2
-0.00030 a^{\dagger 2}a^2b^\dagger b
-0.00634 a^{\dagger 2}a^2
\nonumber\\ & & -2.46444\cdot 10^{-7}  a^\dagger \,ab^{\dagger 4}b^4
-6.97206\cdot 10^{-6}  a^\dagger \,ab^{\dagger 3}b^3
-0.00022 a^\dagger ab^{\dagger 2}b^2
\nonumber\\ & & -0.01121 a^\dagger ab^\dagger b
+1.28264 a^\dagger a
+3.44234\cdot 10^{-9}  b^{\dagger 5}\,b^5
\nonumber\\ & & -2.84772\cdot 10^{-7}  b^{\dagger 4}\,b^4
-1.59258\cdot 10^{-5}  b^{\dagger 3}\,b^3
-0.00171 b^{\dagger 2}b^2
\nonumber\\ & & +0.69148 b^\dagger b
+0.99552
\end{eqnarray}

\begin{table}[!ht]
\parbox[b]{.9\textwidth}{\caption[K]{\label{t_one}\bf Comparison between
flow equation calculations and numerical data in a case of
incommensurate frequencies ($w\!=\!1.3, v\!=\!0.7, \lambda\!=\!-0.1,
n\!=\!0.1$)}\vspace{2mm}}
\begin{tabular}{|c|c|c|c|c|c|c|c|c|}\hline
 & & & & & & & & \\
n & $n_1$ & $n_2$ & Cut-off & Iterat. 4 & Iterat. 8 & Numerical & $E_{free}$ &
$\Delta (\%)$ \\[3mm]
\hline
1 & 0 & 0 & 0.995567 & 0.995521 & 0.995525 & 0.995519 & 1.00 & 0.133899 \\
2 & 0 & 1 & 1.687242 & 1.687013 & 1.687010 & 1.686994 & 1.70 & 0.123020 \\
3 & 1 & 0 & 2.278543 & 2.278179 & 2.278170 & 2.278132 & 2.30 & 0.173770 \\
4 & 0 & 2 & 2.375702 & 2.375106 & 2.375064 & 2.375036 & 2.40 & 0.112162 \\
5 & 1 & 1 & 2.959696 & 2.958536 & 2.958439 & 2.958353 & 3.00 & 0.206497 \\
6 & 0 & 3 & 3.060918 & 3.059734 & 3.059592 & 3.059551 & 3.10 & 0.101362 \\
7 & 2 & 0 & 3.549267 & 3.548183 & 3.548119 & 3.547947 & 3.60 & 0.330432 \\
8 & 1 & 2 & 3.637534 & 3.635141 & 3.634827 & 3.634664 & 3.70 & 0.249480 \\
9 & 0 & 4 & 3.742857 & 3.740832 & 3.740491 & 3.740435 & 3.80 & 0.094015 \\
10 & 2 & 1 & 4.219694 & 4.216860 & 4.216555 & 4.216180 & 4.30 & 0.447387 \\
11 & 1 & 3 & 4.312030 & 4.307926 & 4.307197 & 4.306912 & 4.40 & 0.306162 \\
12 & 0 & 5 & 4.421492 & 4.418334 & 4.417653 & 4.417578 & 4.50 & 0.090995 \\
$\vdots$ & $\vdots$ & $\vdots$ & $\vdots$ & $\vdots$ & $\vdots$ & $\vdots$ &
$\vdots$ & $\vdots$\\
80 & 1 & 14 & 11.502109 & 11.437765 & 11.386826 & 11.348431 & 12.10 & 5.108646
\\
81 & 8 & 1 & 11.496693 & 11.460618 & 11.465128 & 11.412886 & 12.10 & 7.603105
\\
82 & 7 & 3 & 11.535246 & 11.478808 & 11.477820 & 11.415802 & 12.20 & 7.908462
\\
83 & 6 & 5 & 11.591118 & 11.517846 & 11.506570 & 11.432484 & 12.30 & 8.540015
\\
84 & 5 & 7 & 11.664419 & 11.578700 & 11.552905 & 11.470273 & 12.40 & 8.887770
\\
85 & 0 & 16 & 11.659609 & 11.614802 & 11.583460 & 11.532429 & 12.20 & 7.644281
\\
\hline
\end{tabular}
\end{table}
Table \ref{t_one} shows the eigenvalue spectra derived from the
transformed Hamiltonian for the cut--off procedure carried out to
fourth order and for the iteration procedure to fourth and to eighth
order in $\lambda$.  The numerical calculations were performed by
diagonalizing a 900$\times$900--matrix using FORTRAN--routines
presented in \cite{eispack}.  The error $\Delta$ gives the relation of
the deviation from the numerical value with respect to the total shift
due to the coupling:
\begin{equation}
  \Delta = \left|\frac{\,E_{Iter_8} - E_{num.}\,}{\,E_{num.}-
E_{free}\,}\,\right|
\end{equation}
It was calculated for the values of the iteration procedure to 8th
order.

The calculated spectrum reproduces the numerical data well. For the
fixed value of $\lambda=0.1$ the iteration procedure gives better
results than the cut--off procedure.  Iteration to higher orders does
not necessarily improve the approximation of the numerical data: the
ground state energy is more accurate for the iteration to 4th order
than for the iteration to 8th order.  Therefore there is no monotonous
convergence of the determined eigenvalues to the exact ones with
increasing order of iteration.  Whether the eigenvalues determined in
the flow equation procedure are always above the exact ones could not
be shown analytically.

\begin{figure}[!ht]
\begin{center}
  \mbox{
    \epsfig{file=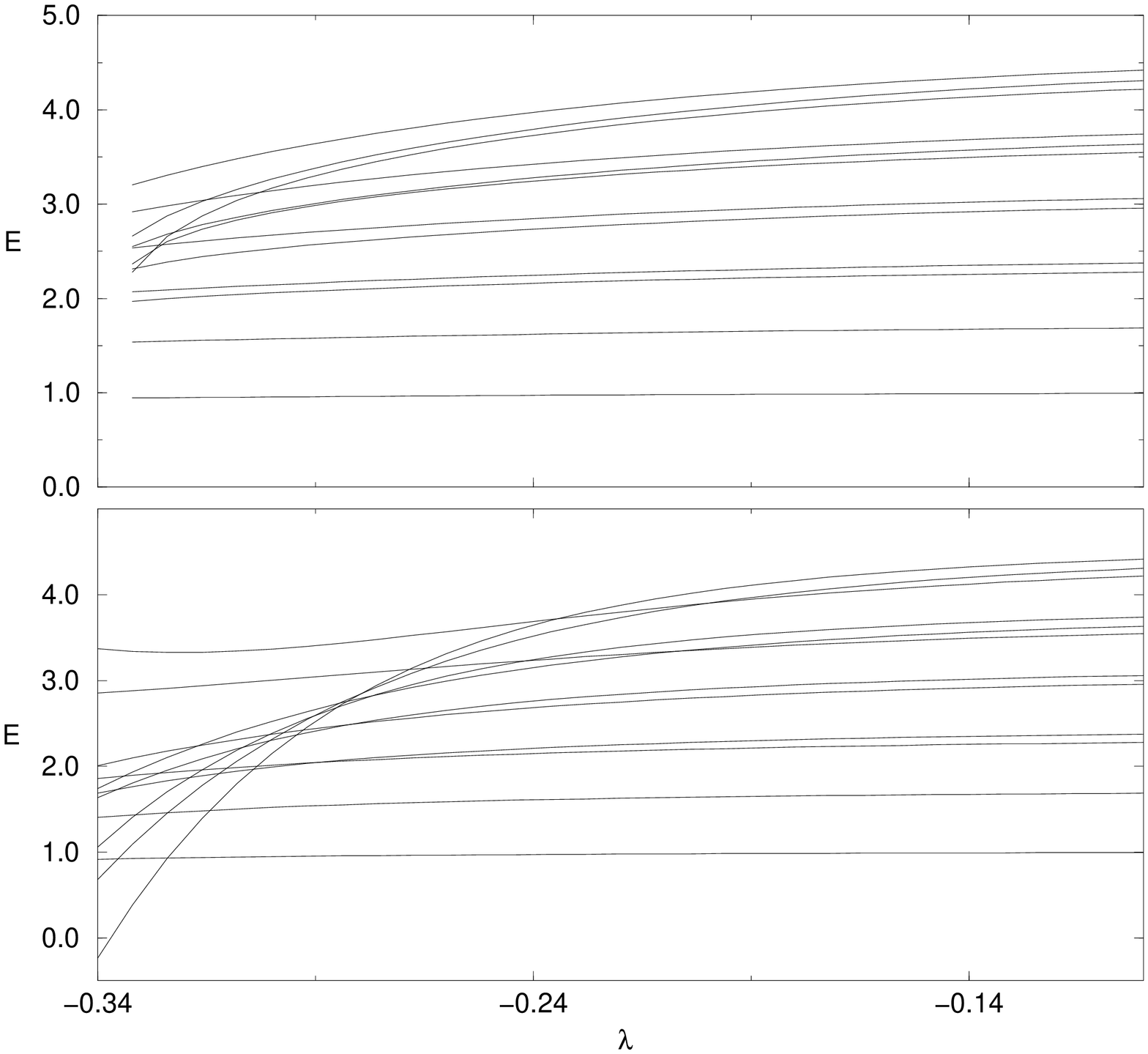,height=0.8\textwidth,angle=270}
    \rule{0.07\textwidth}{0cm}}
  \parbox[b]{.85\picttext}{\caption[K]{\label{comp}The eigenvalues for
      the quantum numbers of the lowest 12 eigenstates at
      $\lambda\!=\!-0.1$ as a function of the coupling strength
      $\lambda$ in the Cut--off procedure to 3rd order (above) and the
      iteration procedure to 8th order (below).  The numerical
      integration breaks down for large coupling (above).}}
\end{center}
\end{figure}
One can determine how the energy eigenvalues for a given pair of
quantum numbers changes as the coupling strength $\lambda$ is
increased.  The result is shown in figure \ref{comp} for 12
eigenvalues in the cut--off procedure to 3rd order and the iteration
procedure to 8th order.  With growing coupling $\lambda$ the potential
well in figure \ref{potential} becomes shallower and the eigenstates
move closer together.  The eigenvalues decrease as the coupling is
increased.  Moreover, one finds that states with higher energy at
$\lambda=-0.1$ will drop faster than the lower states as the coupling
is increased.  This is due to the fact that the eigenstates to higher
eigenvalues are more spread out in space such that the effect of the
$\lambda\cdot q_2^3$--term in the potential (\ref{eq:hh_pq}) upon
them is larger.

Comparison with numerical data seems to indicate that for larger
coupling values of $\lambda$ the cut--off procedure gives more
accurate results than the iteration procedure.  However, a precise
quantitative analysis is not possible in the $\lambda$--range in which
the eigenvalues from the two flow equation procedures differ: for
growing values of $\lambda$ one has to restrict the numerical
diagonali\-zation to smaller matrices in order to avoid the effect
of the continuum causing the appearance of intermittent states (seen
in figure \ref{mess}).  \addtolength{\picttext}{-2cm}
\begin{figure}[!ht]
\begin{center}
  \mbox{
    \epsfig{file=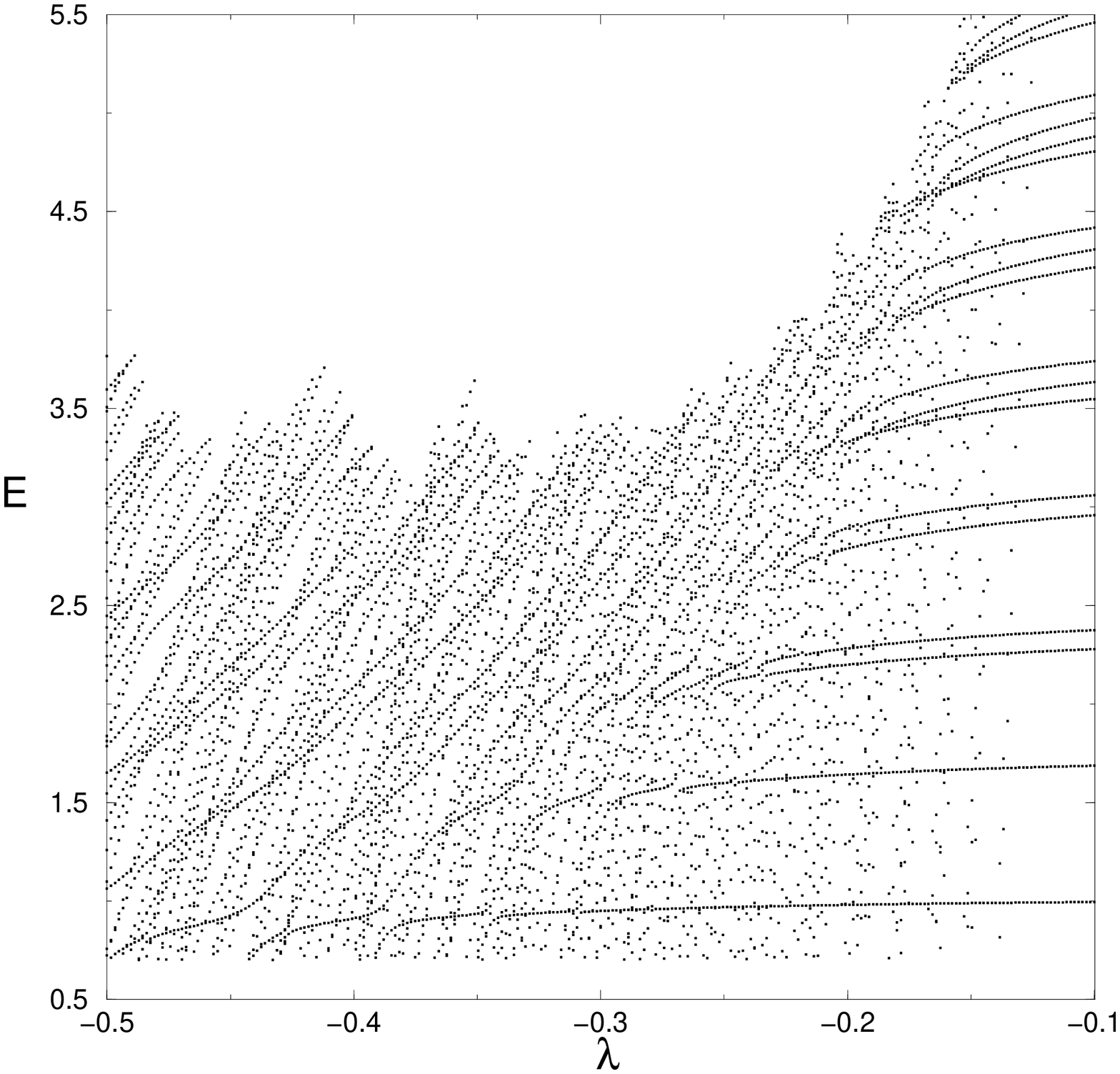,width=8cm,
      height=\textwidth,angle=270}\rule{0.05\textwidth}{0cm}}
  \parbox[b]{\picttext}{\caption[K]{\label{mess}Numerical
      diagonalization of a 900$\times$900--matrix.  For 256
      $\lambda$--values in the interval [-0.5,-0.1] the first 29
      eigenvalues E above 0.7 were determined.  With growing
      perturbation the effect of the continuum becomes dominant.}}
\end{center}
\end{figure}
\addtolength{\picttext}{+2cm}

\Section{A Case of Commensurate Frequencies}
\label{commens}
The two flow equation procedures (sections \ref{cut} and \ref{iter})
were applied to the Hamiltonian (\ref{eq:hh_pq}) in a case of
commensurate frequencies ($w\!=\!1.0, v\!=\!1.0, \lambda\!=\!-0.1,
n\!=\!0.1$). As shown in section \ref{iter}, off-diagonal terms will
not entirely disappear from the transformed Hamiltonian due to the
commensurability.  According to (\ref{eq:repro}) and
(\ref{eq:lambda}) remaining off-diagonal operator terms are of the
form $a^{\dagger k} a^r b^{\dagger m} b^n$ with $\epsilon_{krmn} =
(k-r) + (m-n) = 0$.  They couple states $\ket{n_1,n_2}$ for which
$n_1+n_2=\,$const.  Neglecting these off-diagonal terms, we obtained 
fairly accurate eigenvalues.  To account for the off-diagonal
terms small tridiagonal matrices were numerically diagonalized within
the originally degenerate subspace.  This improves the precision of
the calculated eigenvalues.  The cut--off procedure does not have this
problem. Since the oscillator frequencies $v$ and $w$ in (\ref{eq:hh})
are $\ell$ dependent (see (\ref{eq:dequns})), initially commensurate
frequencies become incommensurate for finite $\ell$. The results for
the cut-off procedure to fourth order, iteration procedures to 4th and
6th order and the improved values of the 6th-order iteration are
listed in table \ref{t_two}.  
\begin{table}[!ht]
\parbox[b]{.9\textwidth}{\caption[K]{\label{t_two}\bf Comparison between
several flow equation calculations and numerical data in a case of commensurate
frequencies \\($w\!=\!1.0, v\!=\!1.0, \lambda\!=\!-0.1,
n\!=\!0.1$)}\vspace{2mm}}
\begin{tabular}{|c|c|c|c|c|c|c|c|c|c|}\hline
 & & & & & & & &  & \\
n & $n_1$ & $n_2$ & Cut-off & Iterat. 4 & Iterat. 6 & improved & Numerical &
$E_{free} $ & $\Delta (\%)$ \\[3mm]
\hline
1 & 0 & 0 & 0.997021 & 0.996989 & 0.996990 & 0.996990 & 0.996987 & 1 & 0.084\\
\hline
2 & 1 & 0 & 1.983815 & 1.983444 & 1.983415 & 1.983415 & 1.983420 & 2 & 0.030\\
3 & 0 & 1 & 1.991400 & 1.991187 & 1.991180 & 1.991180 & 1.991170 & 2 & 0.111\\
\hline
4 & 2 & 0 & 2.962272 & 2.961007 & 2.960816 & 2.956997 & 2.957081 & 3 & 0.195\\
5 & 1 & 1 & 2.968458 & 2.967112 & 2.966985 & 2.966985 & 2.966957 & 3 & 0.084\\
6 & 0 & 2 & 2.985028 & 2.984515 & 2.984469 & 2.988288 & 2.988269 & 3 & 0.162\\
\hline
7 & 3 & 0 & 3.932391 & 3.929514 & 3.928861 & 3.917954 & 3.918277 & 4 & 0.395\\
8 & 2 & 1 & 3.937177 & 3.933515 & 3.933009 & 3.926540 & 3.926527 & 4 & 0.018\\
9 & 1 & 2 & 3.952348 & 3.949675 & 3.949312 & 3.960219 & 3.960177 & 4 & 0.105\\
10 & 0 & 3 & 3.977903 & 3.976969 & 3.976841 & 3.983311 & 3.983255 & 4 & 0.331\\
\hline
11 & 4 & 0 & 4.894173 & 4.888802 & 4.887144 & 4.865409 & 4.865039 & 5 & 0.274\\
12 & 3 & 1 & 4.897560 & 4.890232 & 4.888878 & 4.870960 & 4.871053 & 5 & 0.072\\
13 & 2 & 2 & 4.911331 & 4.904683 & 4.903555 & 4.914437 & 4.916403 & 5 & 2.352\\
14 & 1 & 3 & 4.935486 & 4.931130 & 4.930349 & 4.948267 & 4.948041 & 5 & 0.435\\
15 & 0 & 4 & 4.970025 & 4.968546 & 4.968277 & 4.979130 & 4.978266 & 5 & 3.973\\
\hline
16 & 4 & 1 & 5.849605 & 5.837100 & 5.834141 & 5.796496 & 5.795370 & 6 & 0.550\\
17 & 5 & 0 & 5.847617 & 5.838708 & 5.835184 & 5.799784 & 5.799166 & 6 & 0.308\\[-1mm]
$\vdots$ & $\vdots$ & $\vdots$ & $\vdots$ & $\vdots$ & $\vdots$ & $\vdots$ &
$\vdots$ & $\vdots$ & $\vdots$\\
\hline
\end{tabular}
\end{table}
$\Delta$ gives the error of the improved
values with respect to the total shift due to the coupling:
\begin{equation}
  \Delta = \left|\frac{\,E_{impr.} - E_{num.}\,}{\,E_{num.}-
E_{free}\,}\,\right|
\end{equation}
The originally degenerate subspaces are separated by horizontal lines.

\Section{Dynamics of the Quantum Mechanical System}
\label{dynamics}
In the framework of classical mechanics one can put a particle in the
potential well (figure \ref{potential}) and calculate the trajectory
for a given set of starting values.  
To model this classical approach in the framework of quantum mechanics
one can ask: How does a state originally located within the potential
well evolve with time ?
The eigenstates of the uncoupled oscillator are located within the
potential well.  Their time evolution is given by the Hamilton
operator $H$ of the coupled system.  The absolute value of the matrix
element
\begin{equation}
\label{eq:matrixel}
  \bra{\beta\!\:} \exp(iHt)\!\: \ket{\alpha}
\end{equation}
indicates how much of a particle is in state $\ket{\beta}$ after time
$t$ if the particle was located in state $\ket{\alpha}$ at time 0
where $\ket{\alpha}$ and $\ket{\beta}$ represent two of the
eigenstates of the uncoupled oscillator.

To calculate such matrix elements the eigenstates of the uncoupled
oscillator are expressed in terms of the eigenstates of the full
Hamiltonian of the coupled system.  For this purpose one can set up
flow equations for the transformation of states which results in a
fairly large set of differential equations.  Since the transformation
of the Hamiltonian was already calculated there is an easier way to
calculate the transformation of states which will be explained in the
framework of the iteration procedure introduced in section \ref{iter}:

In analogy to the transformation of the Hamiltonian one determines the
transformed annihilation operator $a(\ell) =
U^\dagger(\ell)\,a\,U(\ell)\,$ from the flow equations
\begin{equation}
   \frac{da(\ell)}{d\ell} = [\eta(\ell),a(\ell)] \,,
\end{equation}
and the generator $\eta$ already known from the calculation of
$H(\ell)$.  One obtains the annihilation operator as a power series in
the coupling constant $\lambda$ similar to the one obtained for the
transformed Hamiltonian (\ref{eq:sum}, \ref{eq:hn}).  By definition
the transformed ground state of the uncoupled oscillator $\ket{0}$ is
given by:
\begin{equation}
  a(\ell=\infty)\, \ket{0} = 0 \quad , \quad b(\ell=\infty)\; \ket{0}  = 0
\quad \mbox{and} \quad \langle 0\ket{0}=1\,.
\end{equation}
These equations can be solved for $\ket{0}$ as a function of the
eigenstates $\ket{n,m}_{\infty}$ of the full Hamiltonian:
\begin{equation}
  \ket{0} = \sum_{n,m,i=0}^{\infty} c_{nm}^{(i)} \lambda^i \,
\ket{n,m}_{\infty} \,.
\end{equation}
One can then construct any exited state $\ket{\alpha} = a^{\dagger
  k}(\infty)\,b^{\dagger m}(\infty)\, \ket{0}$.  In the following the
abbreviations $a\!:=a(\infty)$ and $b\!:=\!b(\infty)$ will be used.

\begin{figure}[!ht]
\begin{center}

  \epsfig{file=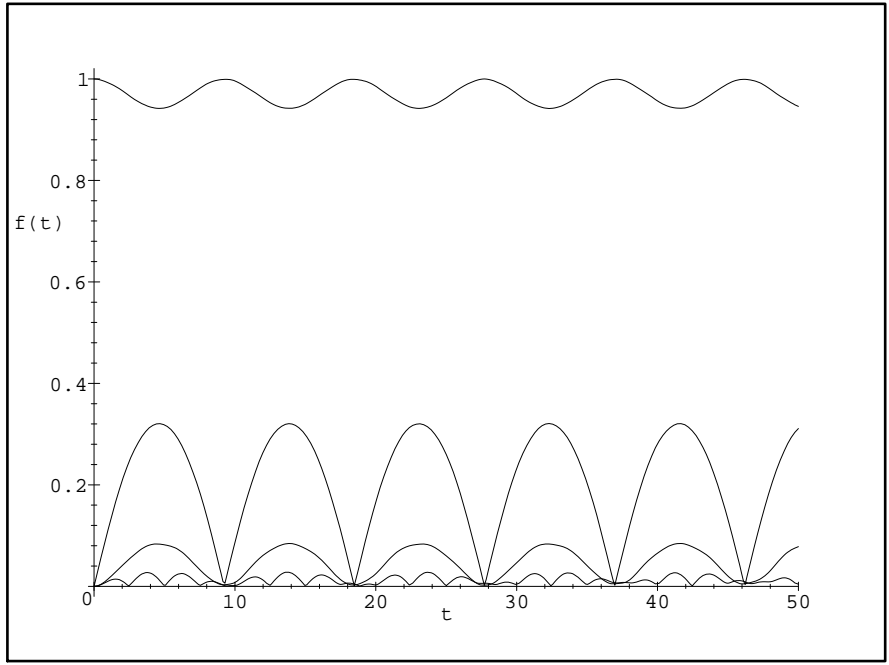,height=0.8\textwidth,angle=270}
  \parbox[b]{.82\picttext}{\caption[K]{\label{ampli1}Transition
      amplitudes $f(t)\! =\! \bra{0}aUa^\dagger\ket{0}$,
      $\bra{0}abUa^\dagger\ket{0}$, $\bra{0}ab^2Ua^\dagger\ket{0}$ and
      $\bra{0}a^3Ua^\dagger\ket{0}$ (in order of size) where\\
      $U\!:=\exp(iHt)$, $a\!=\!a(\infty)$ and $b\!=\!b(\infty)$}}
  \vspace{-0.7cm}
\end{center}
\end{figure}
For several final states $\ket{\beta\,}$ matrix elements of the type
(\ref{eq:matrixel}) are determined as a function time.
Because of the coupling of time and energy in the exponential function
an expansion of the exponent in powers of $\lambda$ -- and thus in
powers of $t$ -- does not give the right long term behavior.
Therefore the exponential functions are not expanded.  The resulting
matrix elements are of the form:
\begin{equation}
  \bra{\beta\!\:} \exp(iHt)\!\: \ket{\alpha} = \!\sum_{k}
a_k\lambda^{b_k}\,e^{iE_kt}\,,
\end{equation}
with coefficients $a_k$ and exponents $b_k$.  $E_k$ represents a
difference of energy eigenvalues which are determined in the iterative
process to the 6th power of $\lambda$.  Thus if the coefficients of
the expansion in $\lambda$ are all of order 1 the phases $E_kt$ can be
determined up to 1 percent accuracy in the range
\begin{equation}
    t<t_{max}\approx \frac{0.01}{\lambda^7} = 10^5\,.
\end{equation}
This gives only a rough estimate since Siegel \cite{Sieg41} proved
that the Birkhoff normal form will generally not converge.  This can
easily be deduced from the fact that the H\'enon--Heiles potential is
not integrable whereas any polynomial in Birkhoff normal form will
always be integrable.  One expects an asymptotic convergence of the
expansion in $\lambda$ such that the coefficients of higher powers of
$\lambda$ will generally grow.  Thus the time range in which the
calculated matrix elements are valid is smaller than the one given
above.

For the initial state $\ket{\alpha}=a^\dagger\ket{0}$ amplitudes for
the transition to the six final states $\ket{\beta} =
a^\dagger\ket{0}$, $a^\dagger b^\dagger\ket{0}$, $a^\dagger b^{\dagger
  2}\ket{0}$, $a^{\dagger 3}\ket{0}$, $a^{\dagger 3}b^\dagger \ket{0}$
and $a^\dagger b^{\dagger 3}\ket{0}$ were determined at $\lambda=-0.1$.
Figure \ref{ampli1} shows the first four of them as a function of time.  
Amplitudes for transitions to final states with an even number of 
$a^\dagger$--$\,$operators can be shown to vanish.

\begin{figure}[!ht]
\begin{center}
  \epsfig{file=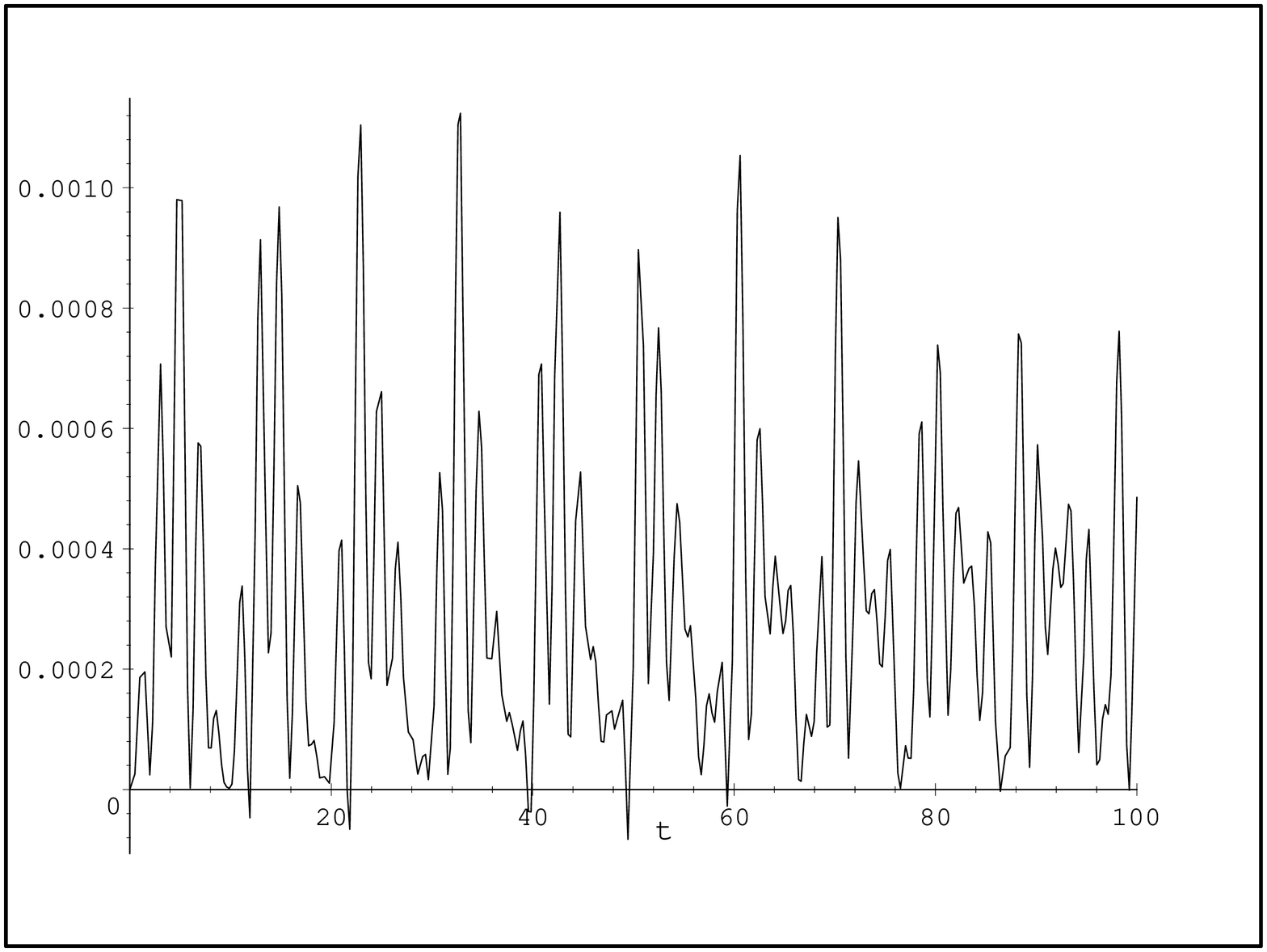,width=0.5\textwidth,
    height=0.75\textwidth,angle=270}\\[-1 ex]
  \parbox[b]{0.75\picttext}{\caption[K]{\label{ampli2}
      $\left(1-\sum_{\beta} |\bra{\beta} U \ket{\alpha}|^2\right)$
      with initial state $\ket{\alpha}\! = \!  a^\dagger \ket{0}$ and
      the six final states $\ket{\beta\!\:}$ mentioned above
      ($U\!:=\exp(iHt)$)}}
\end{center}
\end{figure}

The sum of their absolute values was subtracted from 1 -- in figure
\ref{ampli2} -- to show that these are indeed the relevant
transitions.  The amount of negative value in this figure gives an
indication of the numerical error.

\begin{figure}[ht!]
\begin{center}
  \framebox{\epsfig{file=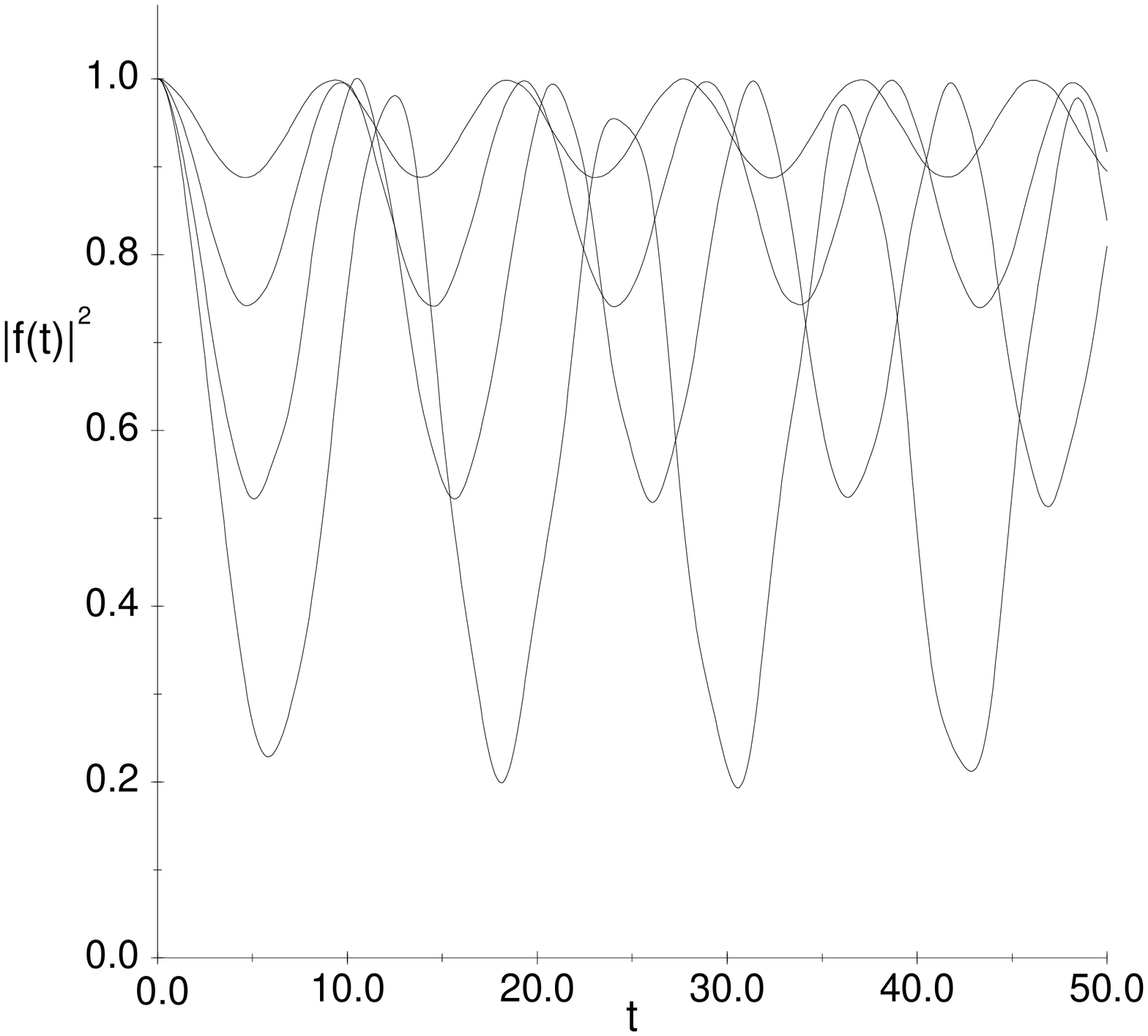,width=0.45\textwidth,
      height=0.7\textwidth,angle=270}}\\[1ex]
  \parbox[b]{0.7\picttext}{\caption[K]{\label{ampli3} Evolution of the
      first exited state $|f(t)|^2$ for the coupling values
      $\lambda\!\!=\!\!-0.1, -0.15, -0.2$ and $-0.25$ (in the order of
      growing amplitude), where $f(t)=\bra{0}a\exp(iHt)\, a^\dagger
      \ket{0}$.}}
\end{center}
\end{figure}

The parameter $\lambda$ is a measure of the strength of coupling
between the two harmonic oscillators (\ref{eq:hh_pq}).  One
expects that a growing coupling value facilitates transitions between
different states.  Figure \ref{ampli3} shows the square of the
transition amplitude $\bra{0}a\exp(iHt)\, a^\dagger \ket{0}$ as a
function of time for various values of $\lambda$.  Moreover, one finds
that the dominant oscillation frequency decreases with growing
coupling.

In the framework of quantum mechanics one expects a particle located
within the potential well of figure \ref{potential} to tunnel through
the potential barrier with a certain probability.  This probability
depends upon the energy of the particle and upon the size and geometry
of the barrier.  
Thus the tunneling should become apparent if one
investigates the dynamics of higher exited states or if one decreases
the size of the potential well by increasing the coupling $\lambda$.
Estimates of the tunneling frequency show that the effect of tunneling
is negligible for $\lambda=-0.1$.  However, for $\lambda=-0.25$
tunneling should become apparent.  
In figure \ref{ampli3} one finds a
modulation in the amplitude of oscillation for $\lambda=-0.25$.  But
this may also be a numerical artifact because in this range of coupling
strength the numerical precision decreases.  Since the Birkhoff
approach consists in approximating a power series by a finite
polynomial, any transition amplitude will always be a finite sum of
harmonic oscillations.  Therefore any calculated time evolution will
necessarily be periodic in time.

Physical observables can be calculated in essentially the same way as
was shown for the transition amplitudes (\ref{eq:matrixel}).
Using the relations
\begin{equation}
\label{eq:pq}
  \hat{q} = \frac{1}{\sqrt{2}}(a^{\dagger}+a) \;, \quad \hat{p}=
\frac{i}{\sqrt{2}}(a^{\dagger}-a) \;.
\end{equation}
for spatial and momentum coordinates one can easily calculate expectation
values of the form $\langle \hat{q}^2 \rangle = \bra{\alpha}U^\dagger
\hat{q}^2 U \ket{\alpha}\,$, which show a similar oscillatory behavior
as that in figure \ref{ampli1}.
\Section{Summary and Outlook}

In this paper we calculated effective eigenvalues of the quantum
mechanical H\'enon--Heiles Hamiltonian using flow equations -- a method
of continuous unitary transformation proposed by Wegner \cite{Wegner94}.  
We used two different procedures to solve the flow equations -- an iterative
procedure and a Cut-off procedure.  The Cut-off procedure has been
used before in several applications of the flow equations. It seems to
be most appropriate if a (perturbative) renormalization of the
Hamiltonian has to be done. This is not necessary in the present case.
The Cut-off procedure has the disadvantage that in most cases it is
difficult to solve the resulting differential equations explicitly.
Often one can extract the asymptotic behavior and based thereon an
approximate solution. But if one wants to have precise numerical
results, one has to solve the differential equations numerically. The
main advantage of the iterative procedure is that the differential
equations can be solved explicitly and that it can be carried out to
much higher orders.

We used both methods to calculate the eigenvalue spectrum of the
H\'enon-Heiles Hamiltonian.  In a case of incommensurate frequencies 
the eigenvalues coincide well with those
obtained by numerical diagonalization of a finite matrix. In the
treated case the precision of results is comparable to that obtained
by Delos and Swimm \cite{DS79} who used consecutive canonical
transformations to approximate the classical Hamilton function by a
Birkhoff normal form.  For small coupling value the iteration
procedure gives more accurate results whereas for larger coupling
values the cut-off procedure seems to be better.  However, for very
large coupling the potential well decreases in size and the discrete
spectrum disappears.  This is found in a series of numerical matrix
diagonalizations at various coupling values.

The case of commensurate frequencies can be treated in essentially the
same way, matrix diagonalization shows good agreement with numerical
results.  In the commensurate
case the diagonalization procedure is much less lengthy than that
presented by Delos and Swimm.  A quantitative comparison to their
results is not possible since the normal form cited in \cite{DS79}
does not correspond to the given parameter set.  And the presented
eigenvalue spectrum does not match with either the parameter set given
or the normal form cited.

On the basis of the transformation of eigenstates we analyzed the
quantum mechanical dynamics of the H\'enon--Heiles system.
We determined the time evolution of a set of states located within the
potential well.
We found oscillations between different states with amplitudes 
depending on the coupling strength.
A certain completeness is found by adding up absolute values of
several transition amplitudes.
 
It is interesting to see that continuous unitary transformations
can be used to obtain precise results for a system like the
quantum mechanical H\'enon--Heiles model. The method was originally
designed to calculate eigenvalues of a given Hamiltonian, at
least approximatively and close to the ground state \cite{Wegner94}.
In many applications of the method, the goal was to obtain
an effective Hamiltonian that describes well the low energy
behavior of the system (see e.g. \cite{subohmic,correl,Lenz,am1}).
The effective Hamiltonian calculated in the present approach
is in a quantum mechanical Birkhoff normal form, i.e. a polynomial in
the oscillator Hamiltonians $(p^2+q^2)$. It does not 
describe the $q^3$--shape of the H\'enon--Heiles potential for large
absolute values of $q$.  
The energies we calculated using this method are energies of approximately
stationary states. As a consequence one cannot describe the
tunneling through the barrier. This limitation is not imposed by the
flow equation method.  Even numerical diagonalizations will not allow a 
correct description of the 
tunneling.  But for sufficiently large escape
times the results describe the short time behavior of the system
quite well. 

The transformation to a Birkhoff normal form yields an asymptotic series 
because the classical H\'enon--Heiles system shows 
a transition to a chaotic regime. 
In principle the flow equations can be successfully applied to a 
quantum mechanical Hamiltonian
that has a chaotic classical counterpart (see also \cite{am2}).
A first focus of interest for such a system are 
statistical properties of the spectrum \cite{Gutz90}. Flow equations
can be used to calculate precise eigenenergies. The way normal ordering 
is introduced determines the interval
of energy where precision is high. Typically one has to
restrict the number of couplings in the Hamiltonian using
some truncation scheme. The neglected terms are in a normal ordered form.
If one chooses the ground state as the basis for the normal ordering,
the resulting effective Hamiltonian describes the ground
state and low lying excitations quite well. In the present approach
the normal ordering was introduced with respect to the ground state
of the uncoupled system. Therefore we were able to obtain precise
results for energies close to the minimum of the potential. One could also
 choose a normal ordering with respect to
some high energy state in order to calculate eigenenergies close
to this energy scale.  While this approach is not sensible in the case of 
the H\'enon--Heiles system,
because the potential well has only a limited depth, it can be useful for 
other physical systems.

\end{document}